\documentclass[sn-mathphys-num, iicol]{sn-jnl}

\usepackage{graphicx}%
\usepackage{multirow}%
\usepackage{amsmath,amssymb,amsfonts}%
\usepackage{amsthm}%
\usepackage{mathrsfs}%
\usepackage[title]{appendix}%
\usepackage{xcolor}%
\usepackage{textcomp}%
\usepackage{manyfoot}%
\usepackage{booktabs}%
\usepackage{algorithm}%
\usepackage{algorithmicx}%
\usepackage{algpseudocode}%
\usepackage{listings}%

\usepackage{hyperref}
\usepackage{ragged2e}
\usepackage{anyfontsize}
\usepackage{footmisc}
\usepackage{subcaption}
\usepackage[braket, qm]{qcircuit}
\usepackage{makecell}
\usepackage{soul} 
\usepackage{setspace}

\begin{document}

\title[Article Title]{Classification and reconstruction for single-pixel imaging with classical and quantum neural networks}

\author*{\fnm{Sofya} \sur{Manko}} \email{manko.sd17@physics.msu.ru}

\author{\fnm{Dmitry} \sur{Frolovtsev}} \email{frolovcev.dmitriy@physics.msu.ru}

\affil{\orgdiv{Department of General Physics and Wave Processes}, \orgname{Lomonosov Moscow State University}, \city{Moscow}, \country{Russia}}


\abstract{Single-pixel cameras are an effective solution for imaging outside the visible spectrum, where traditional CMOS/CCD cameras have challenges. When combined with machine learning, they can analyze images quickly enough for practical applications. Solving the problem of high-dimensional single-pixel visualization can potentially be accelerated via quantum machine learning, thereby expanding the range of practical problems. In this work, we simulated a single-pixel imaging experiment using Hadamard basis patterns, where images from the MNIST handwritten digit dataset and FashionMNIST items of clothing dataset were used as objects. There were selected 64 measurements with maximum variance (6\% of the number of pixels in the image). We created algorithms for classifying and reconstructing images based on these measurements using classical fully-connected neural networks and parameterized quantum circuits. Classical and quantum classifiers showed the best accuracies of 96\% and 95\% for MNIST and 84\% and 81\% for FashionMNIST, respectively, after 6 training epochs, which is a quite competitive result. In the area of intersection by the number of parameters of the quantum and classical classifiers, the quantum demonstrates results no worse than the classical one, even better by a value of about 1-3\%. Image reconstruction was also demonstrated using classical and quantum neural networks after 10 training epochs; the best structural similarity index measure values were 0.76 and 0.26 for MNIST and 0.73 and 0.22 for FashionMNIST, respectively, which indicates that the problem in such a formulation turned out to be too difficult for quantum neural networks in such a configuration for now.}


\maketitle

\section*{}

\begin{footnotesize}\begin{spacing}{1.1}
\textcolor{gray}{This version of the article has been accepted for publication, after peer review but is not the Version of Record and does not reflect post-acceptance improvements or any corrections. The Version of Record is available online at: \url{http://doi.org/10.1007/s11760-025-03875-5}. Use of this Accepted Version is subject to the publisher’s Accepted Manuscript terms of use \url{https://www.springernature.com/gp/open-research/policies/accepted-manuscript-terms}.}
\end{spacing}\end{footnotesize}

\section{Introduction}\label{sec1}

In single-pixel imaging \cite{single-pixel-review}, the light intensity scattered by an object is measured by a single photodetector (not a matrix as in the usual case). Illuminating an object with structured light with a certain set of masks (patterns) allows us to reconstruct an image from such measurements, so that we can obtain spatial resolution without having it in the detector itself. The image is reconstructed by solving the inverse problem, knowing the measurement value and pattern at which it was obtained. The single-pixel imaging method makes it possible to obtain images at wavelengths outside the visible light range (where no cheap imaging camera, such as a CCD camera, is available), with precise time or depth resolution, and also in turbid environments, which causes greater practical significance of the direction. To analyze “single-pixel images” fast enough for practical applications machine learning is used. Quantum machine learning has the potential to speed up the learning process for large-scale problems, which is of great interest to researchers.

Quantum machine learning (QML) is a promising science direction \cite{QML-review-1, QML-review-2, QML-review-3, QML-review-4}, located at the intersection of quantum physics and computer science, in which machine learning methods are developed and studied that can effectively use the unique features of a quantum computer, such as superposition and entanglement \cite{Nielsen-Chuang}. In addition, the variational (parameterized) quantum circuits used in quantum machine learning are resistant to the noise of quantum processors \cite{VQA, noise-res}, which means that they can potentially find useful applications before other known quantum algorithms, which require a fault-tolerant quantum computer. Quantum computing is developing to solve various problems in fields such as logistics, finance, medicine and machine learning \cite{application-logistics, application-finance, application-medicine, application-ML, use-cases}. This work aims to develop this approach for application to the problem of single-pixel imaging and to compare the obtained results with classical neural networks (NNs), which have shown good results in this area in recent years. 

Owing to the fact that quantum processors currently have a small qubit register size and the simulation has exponential complexity depending on the number of qubits, this work uses rather primitive datasets (with a small number of grayscale pixels) and compares with classical models with a simple architecture. The complete classical analog of the parameterized quantum circuits used in this work is a fully-connected neural network. Due to the architecture similarity, we believe that it would be the most valid to compare these two models.

\section{Related work}\label{sec2}

Single-pixel imaging is a promising and cost-effective imaging technique at wavelengths throughout the electromagnetic range (e.g., X-ray \cite{X-ray-1, X-ray-2, X-ray-3}, terahertz \cite{THz-1, THz-2, THz-3}). Single-pixel imaging is often used in the near-infrared range due to the availability of detectors with good sensitivity and sources operating in this range. This wavelength range is particularly suitable for imaging through scattering media such as fog \cite{IR-1, IR-2}, and has also been used to detect and visualize methane leaks \cite{IR-3}.

For practical use of single-pixel imaging systems, it is necessary to significantly reduce the number of patterns and measurements required to increase the speed of this approach. It has been demonstrated that redundancy in the structure of most natural signals or images can be used for this purpose; images are sparse in an appropriate basis, which means that they have many coefficients close to or equal to zero \cite{undefined-1, undefined-2}.

It is also worth noting that many problems do not require reconstruction of the original object; this is the case in applications such as detection or classification \cite{no-rec}.

Recently, work on single-pixel imaging via machine learning methods has begun to appear. For example, in \cite{ML-video}, using neural networks, a set of patterns that were most effective for an object was constructed, and compression of a set of masks up to 4\% was demonstrated to reconstruct 2D images at video signal speed. This model was later used to obtain 3D images in \cite{ML-lidar}. In \cite{ML-fast}, a neural network was used to develop a small number of patterns to classify and identify very fast moving objects. Thus, single-pixel cameras with neural network image processing are an excellent candidate for many practical applications, such as controlling self-driving cars, night vision, gas sensing and medical diagnostics. In addition, it is possible to use alternative types of computing, such as optical machine learning for image reconstruction \cite{optical-ML}.

Quantum machine learning is a rapidly evolving field that has the potential to revolutionize various areas of computing and has achieved a number of great results. The work \cite{QNN4EO} demonstrated the outperformance of a quantum neural network over a classical one in classifying the Earth Observation dataset (EuroSat) by more than 1\% of accuracy. In \cite{quNit}, quantum neural networks (QNNs) were applied to various datasets via the single-shot training scheme, which allows input samples to be trained in an N-level quantum system; it has exceeded a classical NN with a zero hidden layer. However, when two more hidden layers were added to the architecture, the classical NN surpassed the QNN. 

Like classical neural networks, QNNs have several problems, such as a barren plateau that decreases the gradient as the number of qubits increases \cite{plateau-1, plateau-2}. Quantum convolutional neural networks (QCNNs) help to cope with this problem \cite{qCNN-plateaus}. QCNN has achieved excellent classification accuracy despite having a small number of free parameters, which is noticeably better than CNN models under similar training conditions \cite{qCNN}. Recent studies have also explored hybrid quantum–classical convolutional neural networks (containing classical and quantum layers) and demonstrated the classification of images outperforming classical CNNs \cite{HQ-CNN-class-1, HQ-CNN-class-2, HQ-CNN-class-3}. The concept of quantum generative adversarial networks for image generation was implemented experimentally (using a real quantum setup) \cite{qGAN-1, qGAN-2}.

Thus, by introducing quantum machine learning into a single-pixel imaging task, promising results can be obtained.

\section{Data}\label{sec3}

A simple method to obtain an image using a single-pixel detector is to measure each pixel in turn (raster scanning). However, sequentially measuring information about only one pixel in turn is an inefficient use of light capabilities. A more common scanning strategy is to use a sequence of spatially resolved light patterns and measure the scattered intensity as different patterns illuminate the object. To reduce the number of measurements required for high-quality image reconstruction, a set of orthogonal patterns is used, such as the Hadamard basis.

In this work, a vector of numbers obtained by simulating single-pixel measurements is used as data, a set of Hadamard matrices is used as patterns, and the dataset of images of handwritten digits MNIST (28 $\times$ 28, numbers from 0 to 9) and the dataset of items of clothing images FashionMNIST of the same size are used as objects divided into ten classes. Those datasets consist of a training set of 60 000 images and a test set of 10 000 images, which are pre-split. 

\begin{figure}[!htbp]
\includegraphics[width=\linewidth]{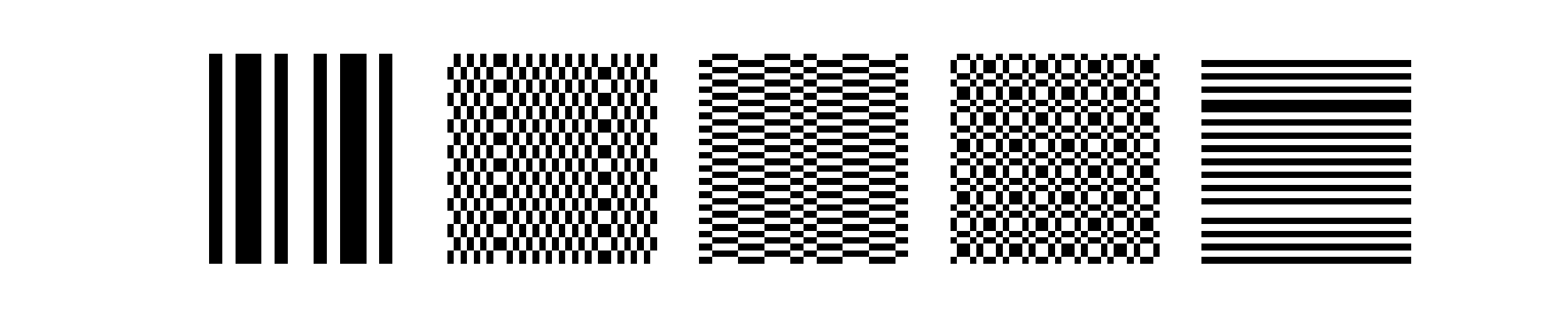}
\caption{\small{Example of several Hadamard patterns $H_n ^{(i)}$) (rows of matrix $H_n$ resized to $n \times n$)}}\label{fig:hadamard}
\end{figure}

Matrix $O$ of size $n \times n$ is our object, which we want to reconstruct, and $M$ is a vector of obtained measurements of length $n^2$. In order to measure the desired object in the Hadamard basis, we need a set of $n^2$ different Hadamard patterns $H_n^{(i)}$ (Fig. \ref{fig:hadamard}), which can be represented as a  matrix:

\vspace{-0.5cm}
\begin{equation} \begin{gathered}
    H_2 =
    \begin{pmatrix}
        1 & 1\\
        1 & -1
	\end{pmatrix}, \\
    H_{n} = 
    \begin{pmatrix}
		H_{n/2} & H_{n/2}\\
		H_{n/2} & -H_{n/2}
    \end{pmatrix} 
    = H_2 \otimes H_{n/2}, 
\end{gathered} \end{equation}
where $n$ is the height and width of the image (object) in pixels. Due to the shape of Hadamard matrices, we expect $n=2^k$. Note that $H_n = H_n^{-1} = H_n^T$, which means that image reconstruction can be performed without matrix inversion.

\vspace{-0.2cm}
\vspace{-0.5cm}

The $i$-th measurement in the Hadamard basis can be represented as a scalar product $M_i = H_n^{(i)} \ast O$, with the pattern and object extended into vectors (row and column, respectively), and the inverse problem of image reconstruction can be represented as $O = \sum \limits _i H_n^{(i)} \times M_i$.

In order to be able to convolve the image with the Hadamard matrix, the dimensions of which are $\varpropto 2^k$, the size of the images was changed to (32 $\times$ 32). For that purpose, we used PyTorch's standard image-resizing function, which is built upon the bilinear interpolation method. Thus, for each object, a measurement vector $M$ of length 1024 (32 $\cdot $ 32) elements can be calculated. 

However, in reality, taking so many measurements (responses to the different light patterns), which are also redundant (since images of objects are usually quite sparse), takes a long time, and their reconstruction is computationally difficult, especially for high-resolution images (with a large number of pixels). However, cameras must capture images at high speed. Therefore, we need to reduce the number of required measurements and leave only the most significant ones for our objects. 

The Hadamard basis decomposition transforms each image from the dataset into 1024 measurements corresponding to each pattern of the orthogonal set. In order to select the most significant patterns (measurements) for the dataset, the variance of each $M_i$ is calculated over the entire training set. The measurements are then sorted by the variance value (importance) and the required number of the most significant ones is selected, these indices ($i$ values) will be the same for all objects in both the training and test sets.

\begin{figure}[!htbp]
\centering
\begin{subfigure}{\linewidth}
\includegraphics[width=\linewidth]{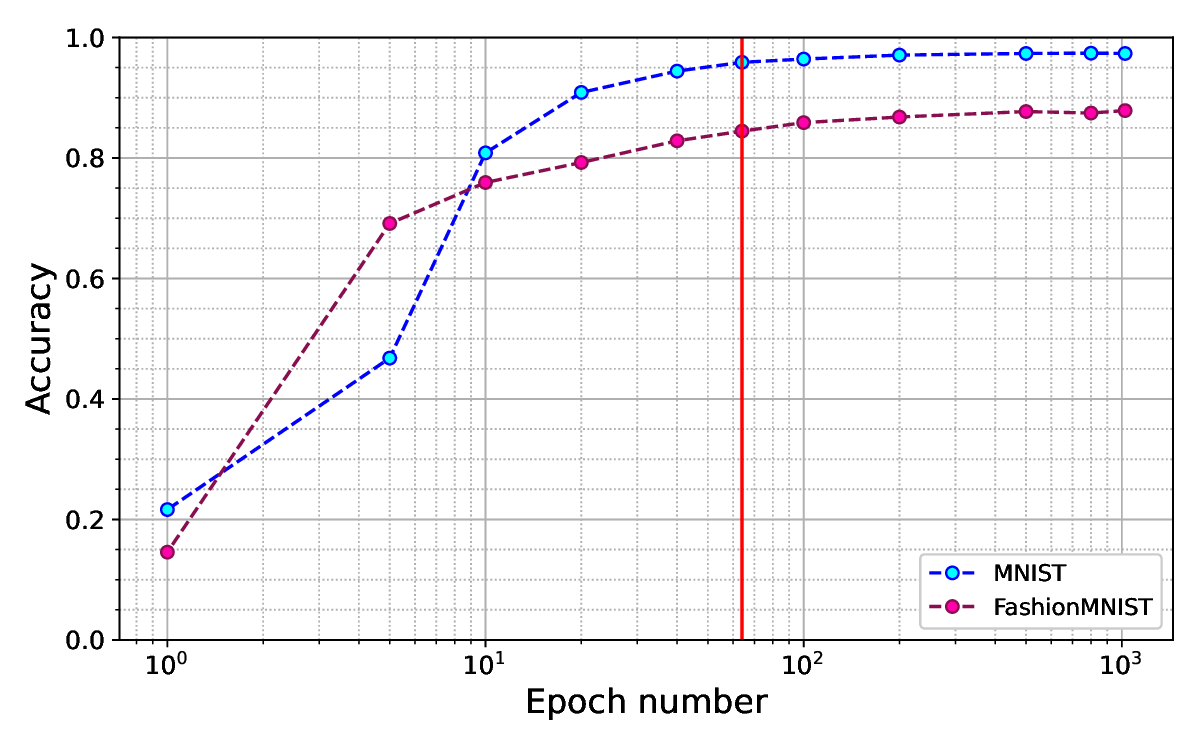}
\caption{}
\label{subfig:acc-on-size}
\end{subfigure}
\begin{subfigure}{\linewidth}
\includegraphics[width=\linewidth]{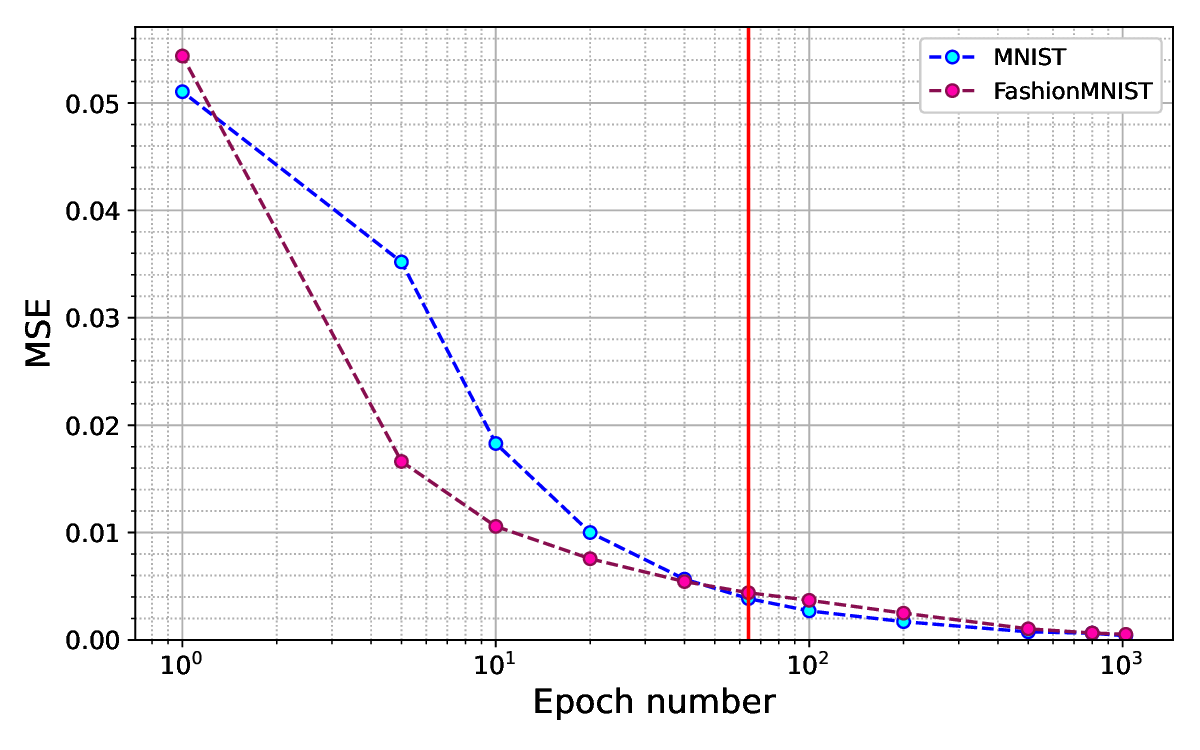}
\caption{}
\label{subfig:mse-on-size}
\end{subfigure}
\caption{Dependence of test set (\subref{subfig:acc-on-size}) - accuracy of classification, (\subref{subfig:mse-on-size}) - mean squared error of image reconstruction, - on the number of measurements in the Hadamard basis in the input layer of the neural network, presented in logarithmic scale on the x-axis}
\label{fig:dependence-on-measurement-number}
\end{figure}

Using classical neural networks, which would be described in section \ref{subsec1}, we investigated the dependence of the training quality on the number of measurements with the greatest variance through the dataset (most significant) in the input layer of the neural network; it is shown in Fig. \ref{fig:dependence-on-measurement-number}. Based on the obtained results, we decided to reduce the number of measurements to 64, which is approximately 6\% of the total number of measurements, making the problem of image reconstruction underdetermined (the number of variables is greater than the number of equations), and one way to solve this problem is neural networks. 

\section{Methods}\label{sec4}

In this section, we detail the architectures of both classical and quantum neural networks for classification and image reconstruction.

\subsection{Classical neural networks}\label{subsec1}

For the classical solution, fully-connected neural networks were used with $ReLu(x)=max(0,x)$ as the activation function, optimized by Adam with a learning rate of 0.0001. The metrics used in this work are cross-entropy \eqref{eq:cross-entropy} loss for the image classification problem and mean squared error (MSE, \eqref{eq:mse}) for the regression problem of image reconstruction. 

The cross-entropy loss function is expressed as follows:
\begin{equation}
    L = -\frac{1}{N}\sum \limits _{i=1} ^{N}\sum \limits _{j=1} ^{M} p_{i,j} \log_{2} q_{i,j},
\label{eq:cross-entropy}
\end{equation}

where $i$ is the number of the element in the batch,
$N$ - their number in the batch, $j$ - class number,
$M$ - number of classes ($M = 10$),
$p_{i}$ - true probability distribution by class for the i-th batch (zero for all classes except the true one under number $j_{\text{true}}$, for which $p_{i,j_{\text{true}}} = 1$),
$q_{i}$ - predicted probability distribution by class for the i-th batch, that is, $q_{3,7}$ is the probability predicted by the neural network with which the 3rd element of the batch belongs to the 7th class (that is, the number “7”). Cross-entropy is a measure of dissimilarity of two distributions. To minimize the loss function, we need to bring the predicted class distribution as close as possible to the true one ($q \rightarrow p$).

The MSE loss function can be written as follows:
\begin{equation}
    L = \frac{1}{N} \sum \limits _{i=1} ^{N} \sum \limits _{j=1} ^{M} (y_{i,j} - f_{j}(x_i)) ^ 2,
\label{eq:mse}
\end{equation}
where $i$ is the number of the element in the batch,
$N$ - their number in the batch,
$j$ - pixel number,
$M$ - number of pixels (in our case $M = 1024$),
$y_{i,j}$ - true value of the $j$-th pixel in the $i$-th batch,
$f_j(x_i)$ is the value of the $j$-th pixel predicted by the model that received a vector of values in the Hadamard basis $x_i$ as input. Here MSE is a metric that characterizes the difference between the true value of a pixel and the one reconstructed by the neural network.

The architecture of the classifier consists of one hidden layer ($64 \rightarrow n_h \rightarrow 10$ neurons for $input \rightarrow hidden \rightarrow output$ layers respectively, results were obtained with different $n_h$). The NN receives 64 of the most important measurements in the Hadamard basis, and the output is 10 numbers characterizing the probabilities of belonging to each of the classes. To reconstruct images, a neural network of one to four hidden layers is used ($64 \rightarrow n_{h_2} \rightarrow ... \rightarrow n_{h_{m-1}} \rightarrow 1024$ neurons). This architecture of the reconstructing NN allows to obtain an output image of 1024 pixels from 64 input measurements.

The expression for calculating the number of parameters of a fully-connected neural network is as follows:
\begin{equation}
n_{\text{params}}^{\text{(cl)}} = \sum \limits _{l = 1} ^ {m} (n_{\text{out}} ^ {(l)} \cdot (n_{\text{in}} ^ {(l)} + 1)),
\label{eq:param-num-cl}
\end{equation}
where $m$ is the number of layers, 
$n_{\text{in}} ^ {(l)}$ - number of neurons at the input of the $l$-th layer,
$n_{\text{out}} ^ {(l)}$ - number of neurons at the output of the $l$-th layer.

\subsection{Quantum neural networks}\label{subsec2}

Quantum neural networks operate with qubits — quantum bits of information. A qubit can be described by a state vector $\ket{\psi} = \cos{\frac{\theta}{2}} \ket{0} + e^{\mathfrak{i} \phi} \sin{\frac{\theta}{2}} \ket{1} = \begin{pmatrix} \cos{\frac{\theta}{2}} \\ e^{\mathfrak{i}\phi} \sin{\frac{\theta}{2}} \end{pmatrix}$, where $\theta, \phi$ are angles on a Bloch sphere \cite{Nielsen-Chuang}. 

To train quantum neural networks, parameterized quantum circuits with a classical optimizer feedback loop are used. A variational quantum circuit is a sequence of quantum gates depending on tunable parameters (rotation operator as in \eqref{eq:Ry}) and entangling gates, which connect qubits and are responsible for nonlinearity (like an activation function in classic) and “neurons” connection.

For quantum solution, parameterized quantum circuits were used with Adam optimizer with learning rate of 0.001 for classification and 0.01 for image reconstruction. For the quantum classification problem, we used Margin loss (\eqref{eq:margin-loss} due to the architectural feature) and MSE \eqref{eq:mse} for image reconstruction.

Margin loss can be described by the following formula:
\begin{equation}
    L = \frac{1}{N} \sum \limits _{i = 1} ^ N \sum \limits _{j \neq y_i} ^ M \max (0, s_j - s_{y_i} + \Delta),
\label{eq:margin-loss}
\end{equation}
where $i$ is the number of the element in the batch,
$N$ - their number in the batch, $j$ - class number,
$M$ - number of classes ($M = 10$),
$s_j$ - the result of the prediction of the j-th classifier from the input data $x_j$, which lies in the interval [-1, 1],
$s_{y_i}$ - true class label for $x_j$ (``1'' or ``-1'').
Using this loss function, the model learns the correct class for each image to have a score higher than the incorrect classes by some fixed margin $\Delta\geq0$ (we set $\Delta = 0.15$). Thus, we want the distance between input values belonging to the same class to be at least $\Delta$ smaller than the distance between input values from different classes.

Consider how a quantum neural network works. First we need to encode classical data to qubits, which could be done by the procedure called amplitude embedding so that we prepare our qubits in state $\ket{\psi_x} = \sum _{i=1} ^{64} x_i \ket{i}$, where $\sum _{i=1} ^{64} |x_i|^2 = 1$. Quantum state complex amplitudes $x_i$ become equal to normalized classical feature values, which are real values. This allows us to encode features to $\log_2 n_{\text{feat}}$ qubits. It is worth clarifying that this method of data encoding as amplitude embedding is not optimal for near-term hardware installations, since the circuit for amplitude embedding has a depth that grows exponentially with the number of qubits, but is acceptable for simulation, since it requires a significantly smaller number of qubits than other encoding methods, which is very important for classical simulation. Research into how to most efficiently encode classical data into qubits is now actively underway in the global community \cite{q-RAM}.

Next, we operate qubits state with a parameterized circuit consisting of layers called ansatz (represented in Fig. \ref{fig:ansatz}). A quantum circuit contains several successive layers, after which we have a quantum state depending on the parameters. We can somehow measure the qubits and, using the results obtained, adjust the parameters of the variational circuit via a classical optimizer.

\begin{figure}[!htbp]
\centering
\scalebox{0.8} {
\Qcircuit @C=.5em @R=.5em{
    \lstick{\ket{q_0}} & \gate{R_Y(\theta_0)} & \ctrl{1} & \qw & \qw & \qw & \qw & \qw & \qw\\
    \lstick{\ket{q_1}} & \gate{R_Y(\theta_1)} & \targ & \ctrl{1} & \qw & \qw & \qw & \qw & \qw\\
    \lstick{\ket{q_2}} & \gate{R_Y(\theta_2)} & \qw & \targ & \ctrl{1} & \qw & \qw & \qw & \qw\\
    \lstick{\ket{q_3}} & \gate{R_Y(\theta_3)} & \qw & \qw & \targ & \ctrl{1} & \qw & \qw & \qw\\
    \lstick{\ket{q_4}} & \gate{R_Y(\theta_4)} & \qw & \qw & \qw & \targ & \ctrl{1} & \qw & \qw\\
    \lstick{\ket{q_5}} & \gate{R_Y(\theta_5)} & \qw & \qw & \qw & \qw & \targ & \qw & \qw\\
\\\\}}
\caption{One layer structure of parameterized quantum circuit}
\label{fig:ansatz}
\end{figure}\vspace{-0.5cm}

Quantum real parameterized rotation gate $Ry$ and entangling gate $CNOT$ used in Fig. \ref{fig:ansatz} can be written as matrices:

\begin{equation}
    R_Y(\theta) = 
    \begin{pmatrix}
    cos(\frac{\theta}{2}) & -sin(\frac{\theta}{2})\\[0.2cm]
    sin(\frac{\theta}{2}) & cos(\frac{\theta}{2})
    \end{pmatrix} 
\label{eq:Ry}
\end{equation}

\vspace{0.2cm}

\begin{equation}
    CNOT = 
    \begin{pmatrix}
    1 & 0 & 0 & 0 \\
    0 & 1 & 0 & 0 \\
    0 & 0 & 0 & 1 \\
    0 & 0 & 1 & 0
    \end{pmatrix}
\label{eq:Cx}
\end{equation}

\vspace{0.2cm}

Now we can describe the architectures of quantum neural networks used in this work.

The classification circuit consists of 6 qubits ($\log_2 64$), highly entangling layers (shown in Fig. \ref{fig:ansatz}) and the measurement of the first qubit, which gives us the expectation value. With bias artificially added, it gives us the prediction result. This circuit is a binary classifier that determines the probability of input data belonging to the selected class or to any other. In other words, it solves a binary problem, where 1 means belonging to the selected class and -1 to some other.

The number of trainable parameters of such a circuit can be expressed as:
\begin{equation}
  n_{\text{params}}^{\text{(bin)}} = n_{\text{qubits}} \cdot (n_{\text{layers}} + 1)\ \text{angles} + 1\ \text{bias},
\label{eq:param-num-q-binary}
\end{equation}
where 1 is added to the number of layers, since after all layers one more without entangling gates is usually inserted.

Margin loss \eqref{eq:margin-loss} allows us to train our networks using a one-against-all strategy. Therefore, we have $n_{\text{classes}}$ different quantum circuits with their own parameters and training independently. 

In total, this quantum classifier has $n_{\text{classes}} \cdot n_{\text{params}}^{\text{(bin)}}$ trainable parameters.
So the number of parameters for quantum classification is as follows:
\begin{equation}
    n_{\text{params}}^{\text{(q-class)}}= n_{\text{classes}} \cdot (n_{\text{qubits}} \cdot (n_{\text{layers}} + 1) + 1)
\label{eq:param-num-q-class}
\end{equation}

The quantum algorithm of “single-pixel images” reconstruction is quite similar to the one described above, but there is no need for different trainable circuits. Another difference is that to produce an image at the output of the circuit, we need 4 more qubits (thus $2 ^ {(6 + 4)} = 1024 = 32 \cdot 32$), which are initially in state \ket{0000}. So our ansatz from Fig. \ref{fig:ansatz} needs to be expanded to 10 qubits. At the end of the circuit, we measure the probabilities of the qubit system being in each possible state ($2^{10}$ numbers). The number of trainable parameters in this quantum network is as follows:
\begin{equation}
  n_{\text{params}}^{\text{(q-rec)}}=n_{\text{qubits}} \cdot (n_{\text{layers}} + 1)\ \text{angles}
\label{eq:param-num-q-rec}
\end{equation}

Due to the long training time using quantum simulator, we decided to use a small part of the MNIST and FashionMNIST datasets (640 images of the first and second classes for the training set and 128 of these classes for the test set) for the reconstruction problem. 

\section{Results}\label{sec5}

\begin{figure*}[!htbp]\captionsetup[subfigure]{font=footnotesize}
\centering
\begin{subfigure}{0.4\linewidth}
\includegraphics[width=\linewidth]{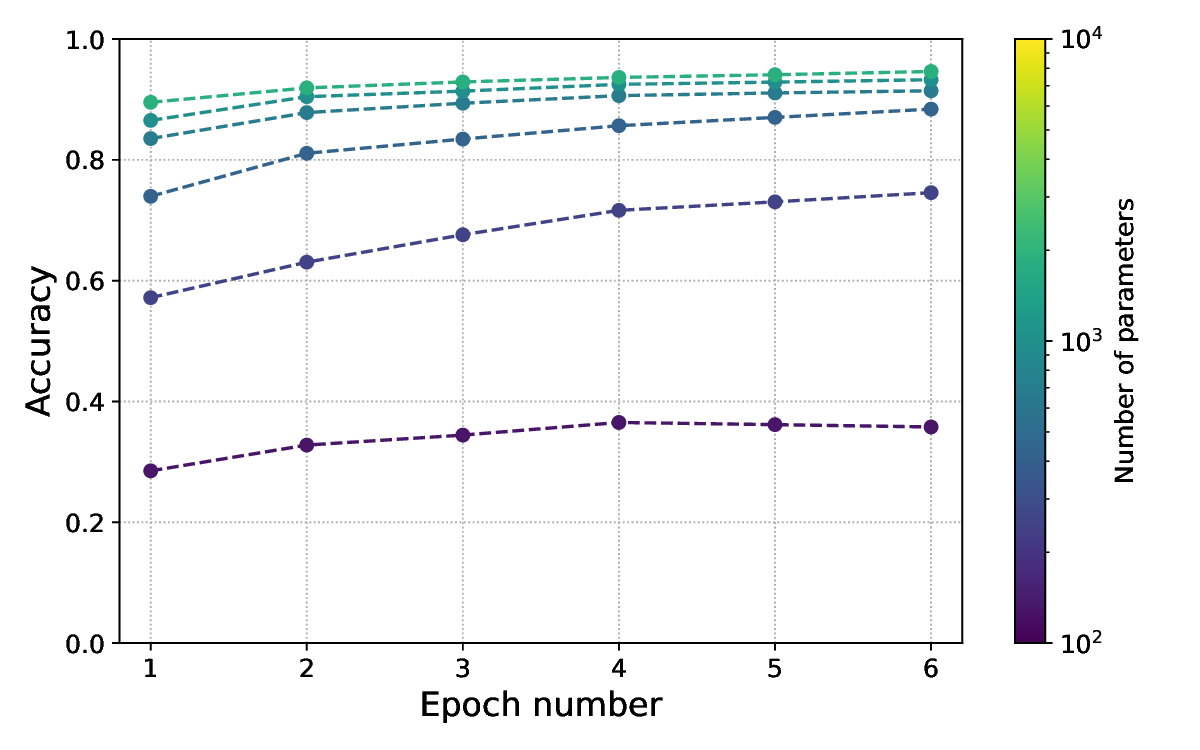}
\caption{quantum MNIST}
\label{subfig:acc-q-mnist}
\end{subfigure}
\begin{subfigure}{0.4\linewidth}
\includegraphics[width=\linewidth]{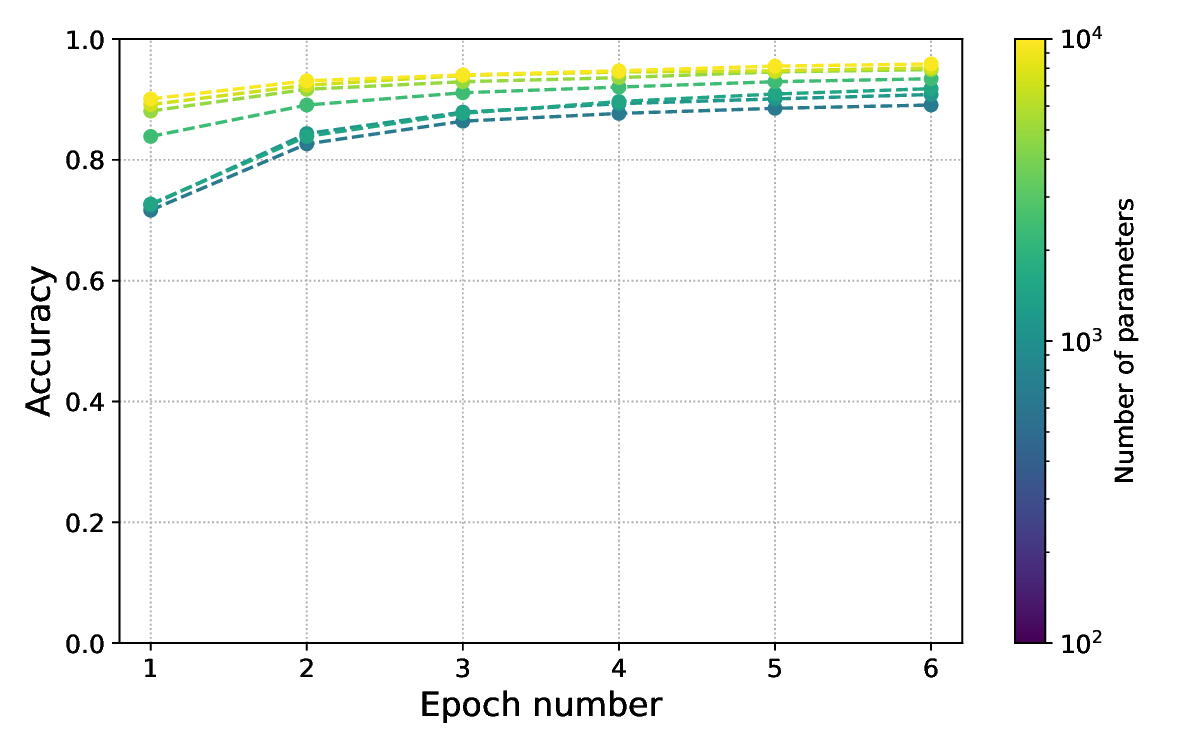}
\caption{classical MNIST}
\label{subfig:acc-cl-mnist}
\end{subfigure}
\begin{subfigure}{0.4\linewidth}
\includegraphics[width=\linewidth]{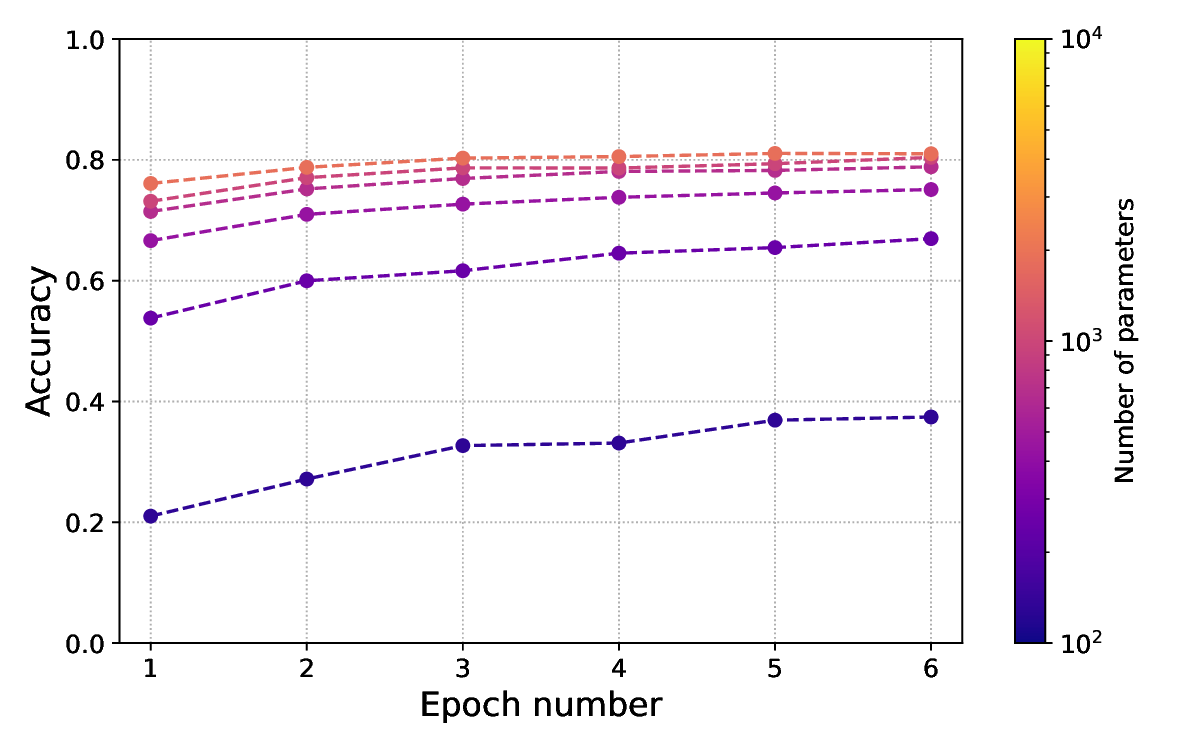}
\caption{quantum FashionMNIST}
\label{subfig:acc-q-fmnist}
\end{subfigure}
\begin{subfigure}{0.4\linewidth}
\includegraphics[width=\linewidth]{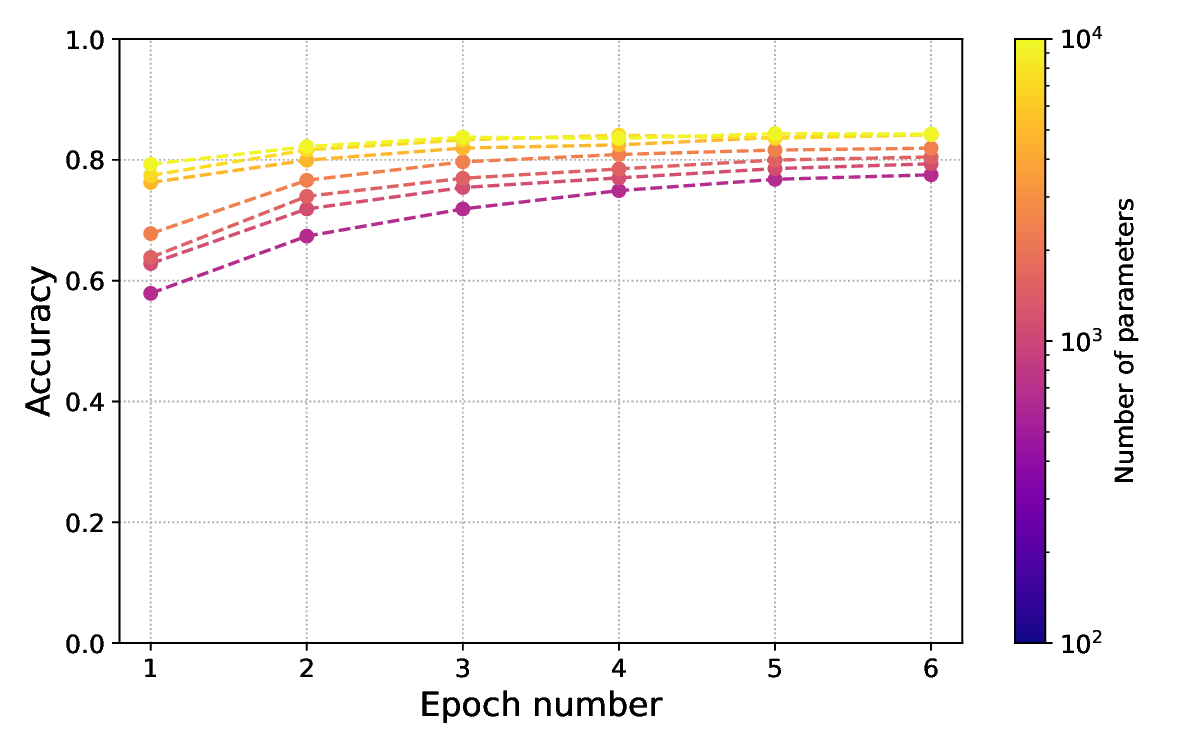}
\caption{classical FashionMNIST}
\label{subfig:acc-cl-fmnist}
\end{subfigure}
\caption{Test accuracy while training various classification neural networks; the number of trainable parameters is shown with a gradient colormap in logarithmic scale}
\label{fig:classification}
\end{figure*}

The neural networks for image classification of MNIST and FashionMNIST datasets by 64 measurements in Hadamard basis presented in the previous section were trained for 6 epochs. We tested NNs with different numbers of trainable parameters, and the training progress over 6 epochs is shown in Fig. \ref{fig:classification}. As the number of parameters increases, the performance of each neural network improves. Figures \ref{subfig:acc-q-mnist} and \ref{subfig:acc-cl-mnist} as well as \ref{subfig:acc-q-fmnist} and \ref{subfig:acc-cl-fmnist} show that quantum and classical classifiers seem to have similar asymptotic behavior. This means that with a sufficient number of parameters \cite{overparam}, a quantum neural network should not be inferior to a classical one.

\begin{figure*}[!htbp]\captionsetup[subfigure]{font=footnotesize}
\centering
\begin{subfigure}{0.4\linewidth}
\includegraphics[width=\linewidth]{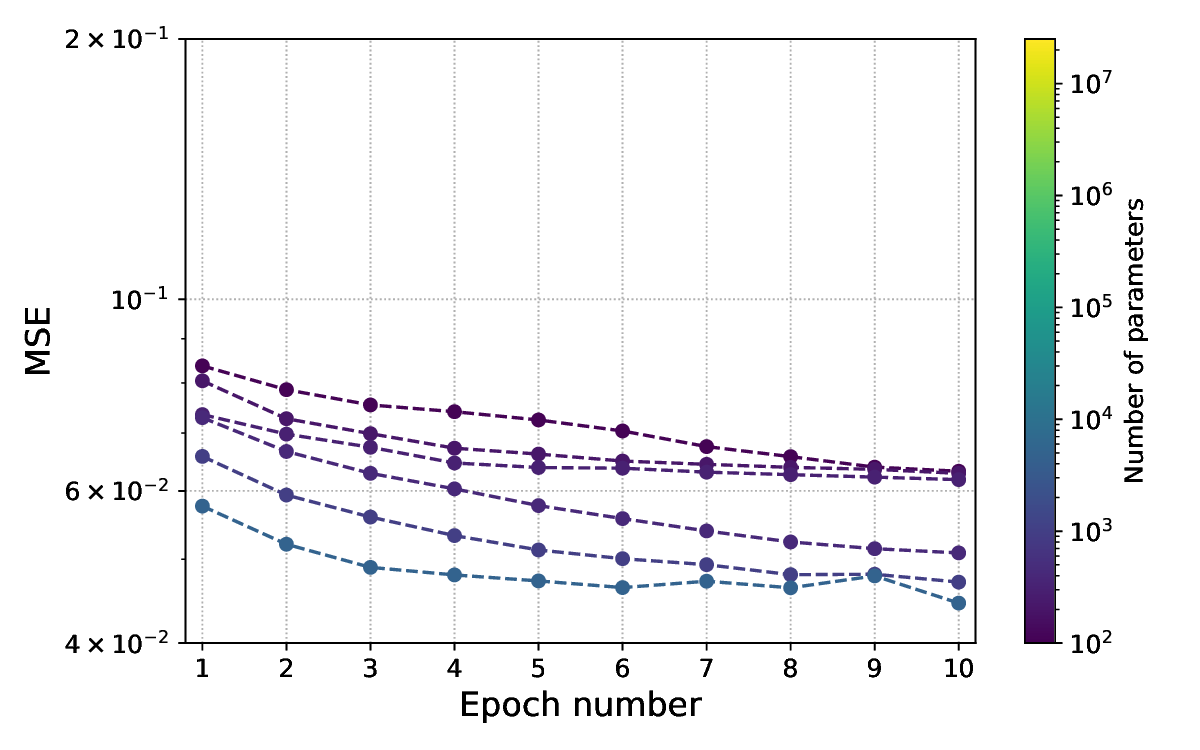}
\caption{quantum MNIST}
\label{subfig:mse-q-mnist}
\end{subfigure}
\begin{subfigure}{0.4\linewidth}
\includegraphics[width=\linewidth]{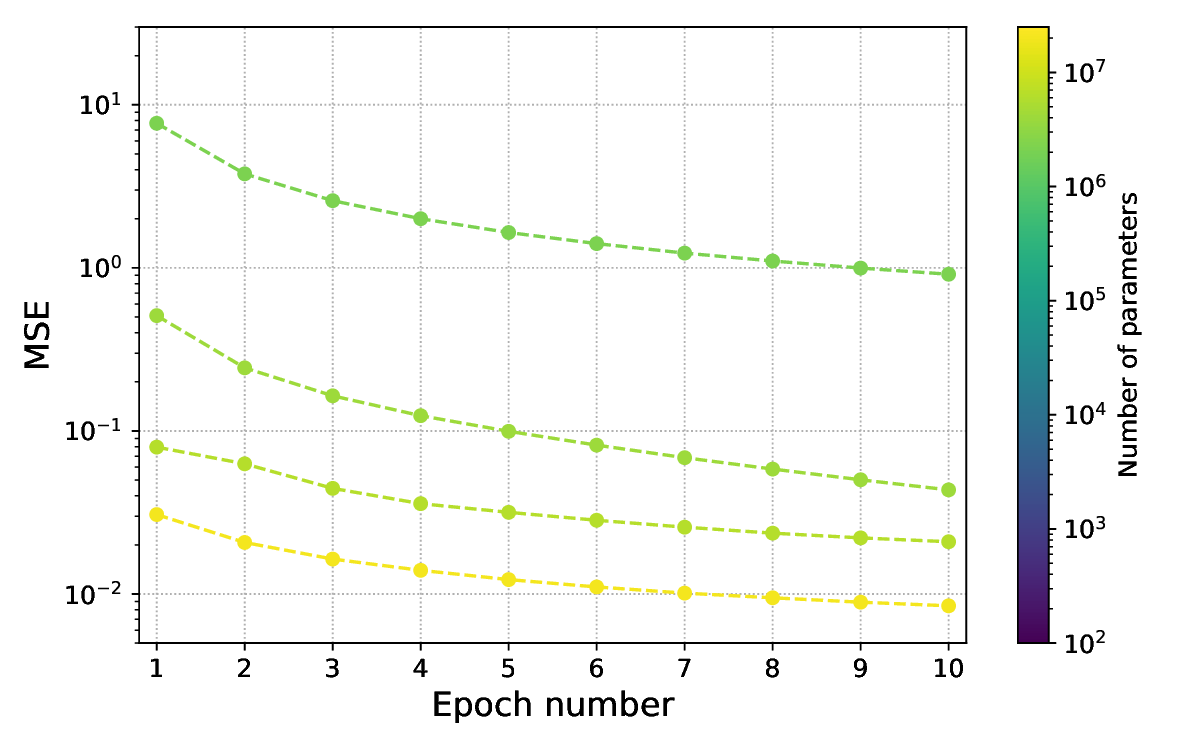}
\caption{classical MNIST}
\label{subfig:mse-cl-mnist}
\end{subfigure}
\begin{subfigure}{0.4\linewidth}
\includegraphics[width=\linewidth]{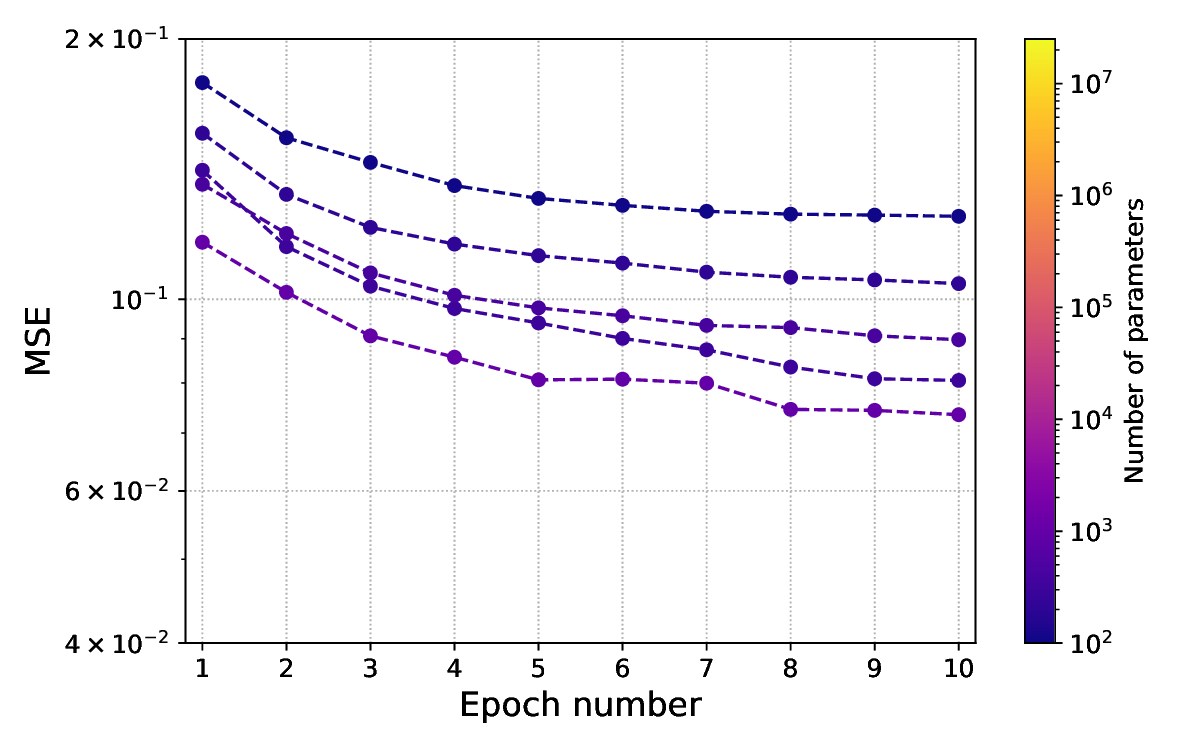}
\caption{quantum FashionMNIST}
\label{subfig:mse-q-fmnist}
\end{subfigure}
\begin{subfigure}{0.4\linewidth}
\includegraphics[width=\linewidth]{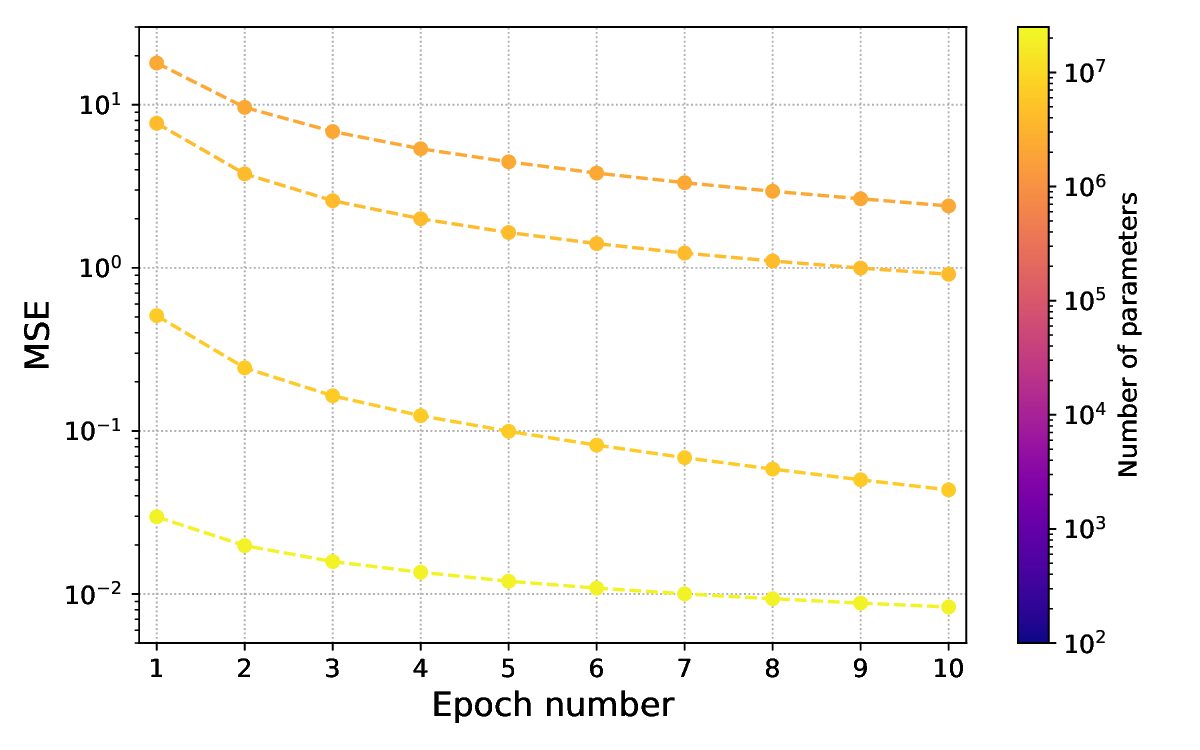}
\caption{classical FashionMNIST}
\label{subfig:mse-cl-fmnist}
\end{subfigure}
\caption{Test MSE while training various reconstruction neural networks, in logarithmic scale on the y-axis; the number of trainable parameters is shown with a gradient colormap in logarithmic scale}
\label{fig:reconstruction}
\end{figure*}

In a classification problem, the convergence asymptote of a NN is determined by the complexity of the dataset and the ability of the model to find patterns needed to distinguish objects. This ability can be improved by using, for example, convolutional neural networks.

\begin{figure}[!htbp]
\includegraphics[width=\linewidth]{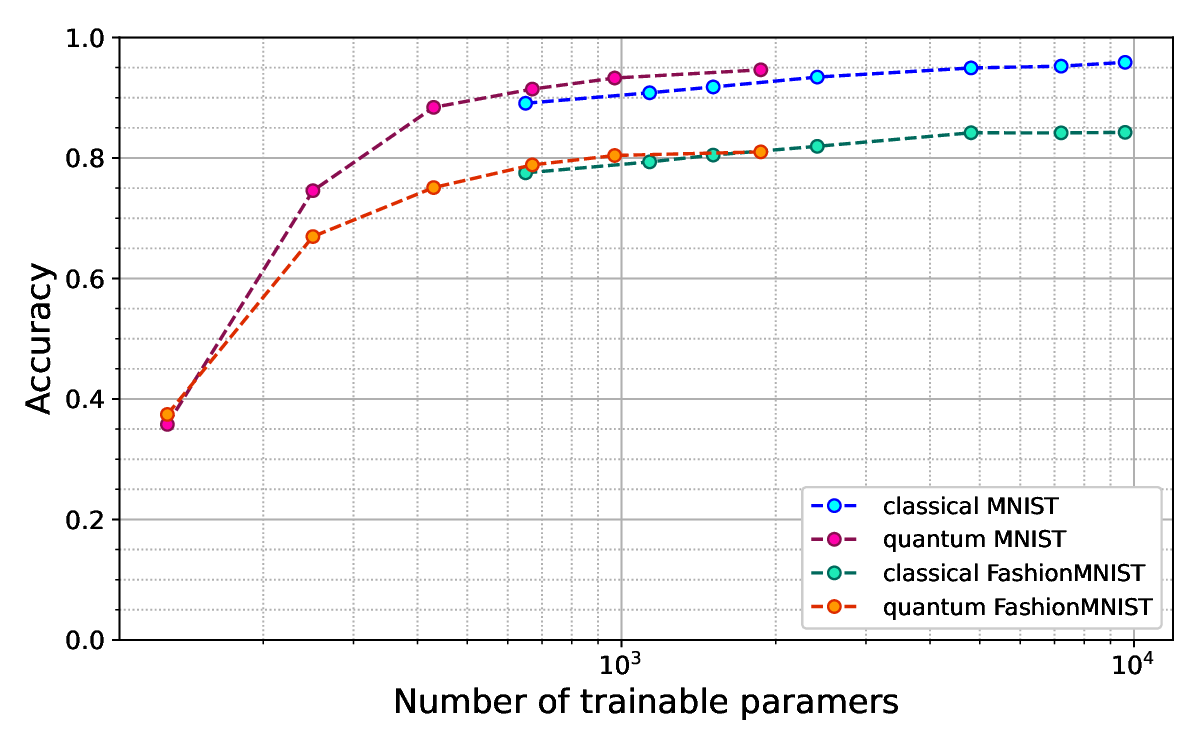}
\caption{Dependence of the quality of models training on the number of trained parameters, presented in logarithmic scale on the x-axis}
\label{fig:performance}
\end{figure}

To compare the efficiency of the classical and quantum models with an equal number of parameters, one can plot the dependence of accuracy on the number of trainable parameters of the model, as shown in Fig. \ref{fig:performance}. The possible number of parameters for training the quantum model is severely limited by the time required to simulate a quantum circuit (more details in Appendix \ref{secA1}). Despite the seemingly small size of the problem, training via quantum neural networks takes a significant amount of time. This is due to the fact that simulating quantum circuits using a classical computer is computationally complex and time-consuming \cite{q-sim}. It is obvious that quantum machine learning loses significantly in speed on small-scale problems due to the large setup overheads; however, with increasing scale, an advantage over classical networks is predicted \cite{QML-review-1, QML-review-2, QML-review-3, QML-review-4}. Nevertheless, as can be seen from the graph, in the area of intersection by the number of parameters of the quantum and classical NNs, the quantum demonstrates results no worse than the classical one, even better by a value of about 1-3\%. Such a result demonstrates not only the applicability of quantum neural networks for the task of classifying "single-pixel images", but also the possibility of potential improvement in the quality of the solution via QNN.

Moreover, the size of the quantum classifier allows it to be run on a real quantum computer.

\begin{figure}[!htbp]\captionsetup[subfigure]{font=footnotesize}
\begin{subfigure}{0.49\linewidth}
\includegraphics[width=\linewidth]{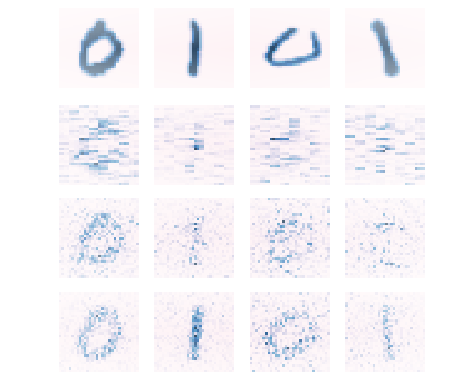}
\footnotesize{\caption{quantum MNIST}}
\label{subfig:rec-q-mnist}
\end{subfigure}
\begin{subfigure}{0.49\linewidth}
\includegraphics[width=\linewidth]{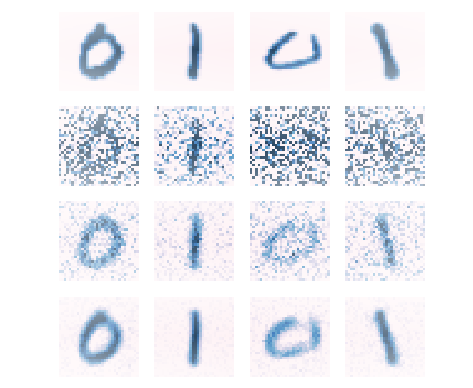}
\caption{classical MNIST}
\label{subfig:rec-cl-mnist}
\end{subfigure}
\begin{subfigure}{0.49\linewidth}
\includegraphics[width=\linewidth]{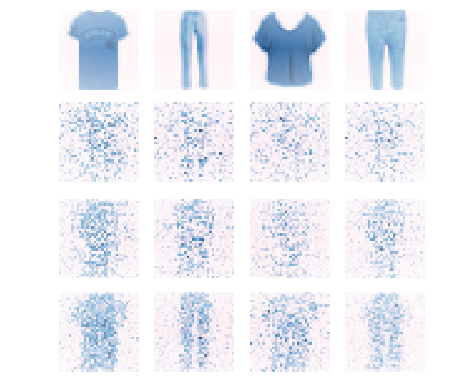}
\caption{quantum FashionMNIST}
\label{subfig:rec-q-fmnist}
\end{subfigure}
\begin{subfigure}{0.49\linewidth}
\includegraphics[width=\linewidth]{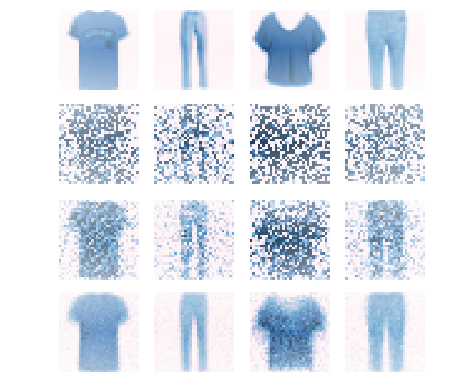}
\caption{classical FashionMNIST}
\label{subfig:rec-cl-fmnist}
\end{subfigure}
\caption{In the first row are the original images (objects). In the rest are images reconstructed by the corresponding networks; from top to bottom: images for the two NNs with the smallest number of trainable parameters and with the largest, which correspond to the two highest and the lowest curves in Fig. \ref{fig:reconstruction}, respectively}
\label{fig:rec-images}
\end{figure}

The progress of NNs training over 10 epochs for image reconstruction in the simulated single-pixel imaging experiment is shown in Fig. \ref{fig:reconstruction}. And the images reconstructed via trained networks with different numbers of parameters are presented in Fig. \ref{fig:rec-images}.

The quality of image reconstruction also improves with an increasing number of parameters. In both cases, quantum and classical neural networks, the reconstructed images change from a ripple of pixels vaguely resembling the original object, for NN with a small number of trainable parameters, to images with a formed object with an increase in the number of parameters.

The gap between the best results for the classical neural network and the quantum one is very large. This is because we quickly reach the computational limit of our computer for simulated such a complex problem, so the difference in the number of trainable parameters between the best results for a quantum and classical neural network is 3 orders of magnitude, which greatly affects the final result. The training time of a quantum neural network depends linearly on the number of layers in it and exponentially on the number of simulated qubits. The dependence of the improvement in the quality of the reconstructed image with an increase in the number of layers does not allow us to hope for the ability to observe a comparable result for a quantum neural network within a reasonable simulation time. In addition, although the required quantum circuit has a small number of qubits and could potentially be implemented in hardware (thereby removing the exponential dependence on the number of qubits and leaving only a linear dependence on the number of layers), unfortunately, the calculation of circuits of such great depth is not available for noisy quantum devices today.

Exact values of training metrics, including training time, can be found in Tables \ref{tab:classification} and \ref{tab:reconstruction} of Appendix \ref{secA1}, as well as thoughts on the possibility of modifying the quantum circuit.

\section{Conclusion}\label{sec6}

Classical and quantum neural networks were built to solve problems of classification and reconstruction of images based on the measurements of an object founded on the MNIST handwritten digits dataset and the FashionMNIST items of clothing dataset in the Hadamard basis reduced to 64 most significant patterns, which is only 6\% of the number of pixels in the original object. 

The constructed classical classifier showed the best prediction accuracy of 96\% for MNIST and 84\% for FashionMNIST. The developed classical neural network, which reconstructs an image in a single-pixel imaging task, made it possible to obtain a high-quality image with a mean-square error of 0.08 and a structural similarity index measure of 0.76 and 0.73 for the MNIST and FashionMNIST datasets, respectively. It was also investigated how the training quality metrics of the created classical neural networks depend on the number of measurements in the Hadamard basis, and it was shown that 64 (6\% of all possible) for such a task is the optimal value.

The quantum machine learning method was used to solve the problem of classifying “single-pixel images”, thereby demonstrating the applicability of quantum neural networks for analysis in the problem of single-pixel visualization. The quantum neural network was created and trained on a quantum simulator, which predicts the class of a “single-pixel image” with the best accuracy of 95\% for MNIST and 81\% for FashionMNIST, which is quite a competitive result and could be even better with a number of parameters comparable to the classical case. In the area of intersection by the number of parameters, the quantum neural network shows classification no worse than the classical one, and even with an accuracy higher by 1-3\%. The quantum neural network for image reconstruction has been developed, and the results of its work have been demonstrated; it reconstructs images with a structural similarity index measure of 0.26 for MNIST and 0.22 for FashionMNIST. Despite the significant difference from images reconstructed via a classical neural network, this result can also be considered successful as proof of concept for now.

The image reconstruction problem turned out to be too complex to be solved with adequate quality using quantum circuits with highly entangling layers. To significantly improve the excellence of the solution, it is necessary to increase the number of parameters by several orders of magnitude, which is impossible to achieve in a reasonable computation time. However, in further research, HQ-CNN may prove to be a more promising architecture for this problem, as was previously demonstrated on a similar problem of image generation \cite{qGAN-1, qGAN-2}. Another way to solve the image reconstruction problem in future studies is to use a dataset of smaller images (e.g., Tetris image dataset \cite{HQ-CNN-class-1} or Pen-Based Recognition of Handwritten Digits \cite{qGAN-1}), which would reduce the depth of the quantum circuit potentially needed to solve the problem. This would open up the possibility of simulating quantum neural networks with image reconstruction quality comparable to classical ones and implementing the algorithm experimentally.

The great flexibility of the quantum approach in the context of machine learning is beyond doubt \cite{q-adv}, so we believe that it is advisable to deepen and expand this study using more powerful simulators and hardware systems to which we currently do not have direct access to search for a quantum advantage in models with an increased number of parameters. For classification hardware implementation, the subject of future research could be focused on finding the most efficient way to encode data into the quantum circuit, as well as the form of the ansatz (type of layers). Since two-qubit (entangling) gates are one of the most important sources of noise in real devices, the most efficient quantum circuit for hardware implementation should contain fewer entangling gates, which means that the number of rotating gates within a layer should be increased to maintain good training performance. The most efficient combination of rotating gates and other quantum circuit modifications requires further research. Since the presented circuits have a relatively small depth, they have a high potential for implementation via promising hardware systems, especially considering the possibility of using a simple zero-noise extrapolation error mitigation technique. There are potentially no restrictions on running the proposed algorithms experimentally. Algorithms of similar complexity have already been demonstrated on available hardware \cite{ExpClass}. However, this task remains the subject of further research due to the low availability of quantum computers.

Single-pixel imaging is a promising and cost-effective imaging technique across the entire electromagnetic spectrum. Combined with recent advances in machine learning algorithms, single-pixel imaging promises to be a powerful method for low-cost, scan-free 3D recognition and classification, which holds promise for critical applications such as object detection and classification, surface mapping, and 3D situation recognition for autonomous vehicles. The results obtained in this work give us hope for new and interesting opportunities that quantum machine learning may provide in the future.

\backmatter

\bmhead{Acknowledgements} 
We would like to thank the entire team of the MSU.AI project for teaching us a lot and giving many important advices, especially Sergey Kolpinskiy. We also express our deep gratitude to the Laboratory of Experimental and Theoretical Quantum Optics, in particular Dmitry Agapov and Anatoly Chirkin, for productive discussions and valuable comments. Furthermore, we would like to acknowledge Alexey Moiseevskiy for crucial recommendations.

\bmhead{Code availability} 
Source code could be found \url{https://github.com/Oounce/classical-quantum-mnist.git}.


\begin{appendices}

\section{Numerical results}\label{secA1}

Table \ref{tab:classification} presents the metrics of the trained classifiers. The classical classifier with 640 parameters is the network without hidden layers, and further reduction of the number of parameters for such an architecture is impossible.
\vspace{-0.5 cm}
\begingroup
\renewcommand*{\thefootnote}{\alph{footnote}}
\begin{table}[!htbp]
\caption{Different classification networks results (for each network, the first row corresponds to MNIST and the second to FashionMNIST)}\label{tab:classification}
\footnotesize
\begin{tabular}{@{}cccc@{}}
\toprule
Architecture\footnotemark  &  $n_{\text{params}}$\footnotemark  &  Accuracy  & 
 Training time \footnotemark  \\
\toprule
\multicolumn{4}{c}{classical} \\\hline\hline
0    &  640    &  \makecell{0.89 \\ 0.78}  &  6.7 s   \\\hline
15   &  1 135  &  \makecell{0.91 \\ 0.79}  &  7.8 s   \\\hline
20   &  1 510  &  \makecell{0.92 \\ 0.80}  &  7.8 s   \\\hline
32   &  2 410  &  \makecell{0.93 \\ 0.82}  &  7.9 s   \\\hline
64   &  4 810  &  \makecell{0.95 \\ 0.84}  &  9.5 s   \\\hline
96   &  7 210  &  \makecell{0.95 \\ 0.84}  &  9.8 s   \\\hline
128  &  9 610  &  \makecell{0.96 \\ 0.84}  &  10.2 s  \\\hline
\multicolumn{4}{c}{quantum} \\
\hline\hline
1    &  130    &  \makecell{0.36 \\ 0.37}  &  5.5 h  \\\hline
3    &  250    &  \makecell{0.75 \\ 0.67}  &  10.0 h  \\\hline
6    &  430    &  \makecell{0.88 \\ 0.75}  &  16.5 h  \\\hline
10   &  670    &  \makecell{0.91 \\ 0.79}  &  24.5 h  \\\hline
15   &  970    &  \makecell{0.93 \\ 0.80}  &  34.5 h  \\\hline
30   &  1 870  &  \makecell{0.95 \\ 0.81}  &  67.0 h  \\\hline
\botrule
\end{tabular}
\footnotetext[a]{The column indicates: for classical neural networks - hidden layer size; for quantum networks - the number of layers (Fig. \ref{fig:ansatz}) in the quantum circuit.}
\footnotetext[b]{According to \eqref{eq:param-num-cl} and \eqref{eq:param-num-q-class} for classical and quantum networks, respectively.}
\footnotetext[c]{Average CPU time for classical solution and PennyLane lightning.qubit simulator time via CPU for quantum solution (CPU: Intel(R) Core(TM) i7-11700 workstation with 64 GB RAM).}
\end{table}
\vspace{-0.3 cm}

It is potentially possible to modify the architecture of the quantum classifier to further reduce the number of trainable parameters by combining ten binary classifiers into one multi-class classifier. The first, most obvious, way is similar to the architecture of a quantum circuit for image reconstruction: one can add 4 additional qubits initialized to state {\ket{0000}}, a 10-qubit ansatz, and a measurement of 10 expected values of the observables. Such an architecture would allow training a quantum classifier using cross-entropy as a loss function. However, increasing the number of qubits exponentially increases the simulation time, which would not allow us to train on the entire dataset (as in the case of reconstruction). Potentially, one can keep a small number of qubits, but this option would require a non-trivial modification of the loss function, which can be the subject of further research.

Table \ref{tab:reconstruction} presents the metrics of the trained NNs reconstructing images in simulated single-pixel experiment.

\vspace{-0.8 cm}
\begin{table}[!htbp]
\caption{Different reconstruction networks results (for each network, the first row corresponds to MNIST and the second to FashionMNIST)}\label{tab:reconstruction}
\footnotesize
\begin{tabular}{@{}ccccc@{}}
\toprule
Architecture\footnotemark  &  $n_{\text{params}}$\footnotemark  &  MSE  &  SSIM\footnotemark  &  Training time \footnotemark \\
\toprule
\multicolumn{5}{c}{classical} \\
\hline
\hline
(2000)                    &  2 179 024  &  \makecell{0.914 \\ 1.379}  & \makecell{0.283 \\ 0.256} &  3.8 s \\
\hline
(1000, 2000)              &  4 166 024  &  \makecell{0.043 \\ 0.099}  & \makecell{0.485 \\ 0.410} &  5.6 s   \\
\hline
\makecell{(1000, 2000, \\ 1000)}        &  6 117 024  &  \makecell{0.021 \\ 0.020}  & \makecell{0.558 \\ 0.565} &  6.4 s   \\
\hline
\makecell{(1000, 2000, \\ 4000, 2000)}  &  20 122 024  &  \makecell{0.008 \\ 0.008}  & \makecell{0.758 \\ 0.725} &  20.2 s   \\
\hline
\multicolumn{5}{c}{quantum} \\
\hline
\hline
10   &  130    &  \makecell{0.063 \\ 0.125}  & \makecell{0.109 \\ 0.069}  &  0.4 h \\
\hline
20   &  250    &  \makecell{0.063 \\ 0.104}  & \makecell{0.104 \\ 0.088}  &  0.8 h \\
\hline
30   &  430    &  \makecell{0.062 \\ 0.081}  & \makecell{0.106 \\ 0.162}  &  1.3 h \\
\hline
40   &  670    &  \makecell{0.051 \\ 0.090}  & \makecell{0.195 \\ 0.148}  &  1.7 h \\
\hline
100  &  1 010  &  \makecell{0.047 \\ 0.074}  & \makecell{0.224 \\ 0.216}  &  6.0 h \\
\hline
500  &  5 010  &  \makecell{0.044 \\ -}  & \makecell{0.257 \\ -} &  107.0 h \\
\botrule
\end{tabular}
\footnotetext[a]{The column indicates: for classical neural networks - hidden layers sizes; for quantum networks - the number of layers in the quantum circuit.}
\footnotetext[b]{According to \eqref{eq:param-num-cl} and \eqref{eq:param-num-q-rec} for classical and quantum networks, respectively.}
\footnotetext[c]{Structural similarity index measure.}
\footnotetext[d]{Average CPU time for classical solution and PennyLane default.qubit simulator time via CPU for quantum solution (CPU: Intel(R) Core(TM) i7-11700 workstation with 64 GB RAM).}
\end{table}
\endgroup
\vspace{-0.3 cm}

For a number of parameters smaller than those indicated in Table \ref{tab:reconstruction}, both quantum and classical neural networks for image reconstruction produce results that are completely different from the original images and therefore do not make sense to compare them. The training time increases linearly with the depth of the quantum circuit. However, as can be seen from the last row of the table, a further increase in the number of parameters in the quantum circuit leads to a lack of RAM (training took more than 100 hours instead of the expected 25 hours). There is a potential method to combat this problem by sequentially training one layer (or group of layers), then the next, and so on until the end.  Such a solution would be suitable for a simulator, but anyway, circuits of such great depth would be unachievable for near-term quantum hardware.

In addition, the authors note that the training time demonstrated in this work is not optimal since not all the possibilities of multiprocessing were used.

\begin{figure}[!htbp]\captionsetup[subfigure]{font=footnotesize}
\begin{subfigure}{\linewidth}
\hspace{-0.7cm}
\begin{minipage}[m]{0.85\linewidth}
    {\includegraphics[width=\linewidth]{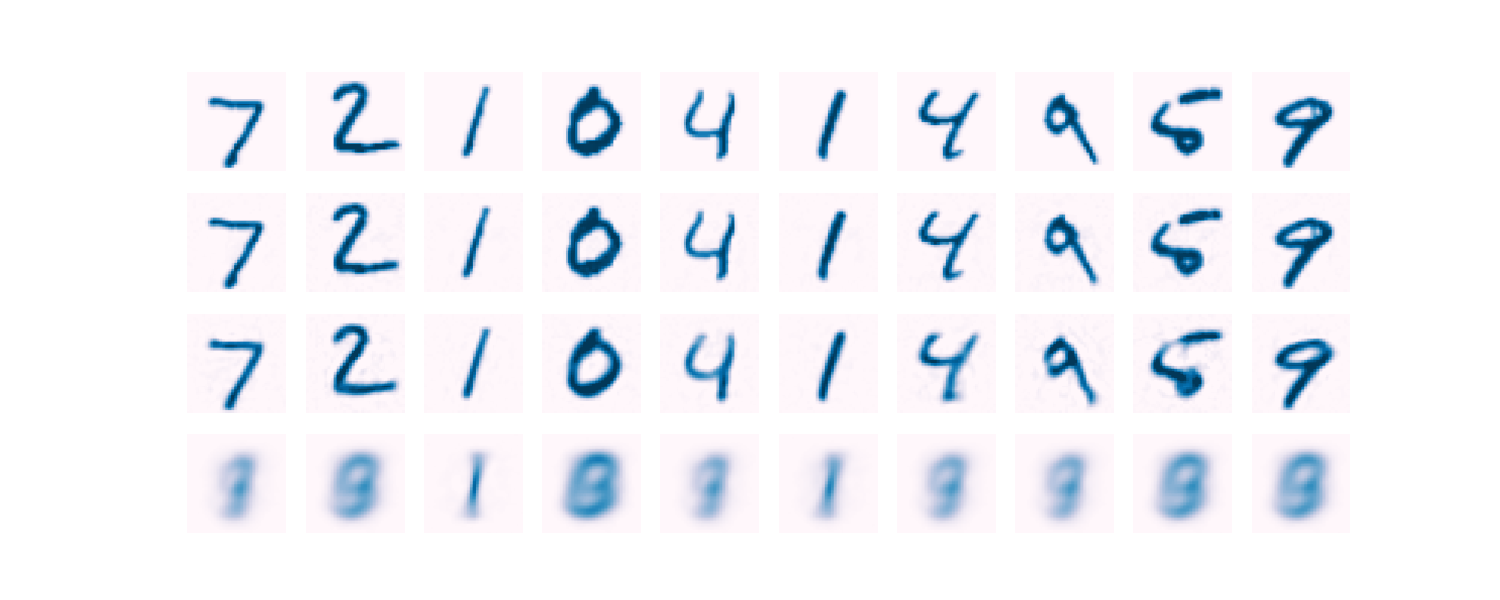}}
\end{minipage}
\hspace{-0.7cm}
\begin{minipage}[m][0.25\linewidth][l]{0.2\linewidth}
    \vfill
    \tiny{original}\vfill
    \tiny{1024\\(SSIM=0.90)}\vfill
    \tiny{64\\(SSIM=0.85)}\vfill
    \tiny{1\\(SSIM=0.17)}\vfill
\end{minipage}
\caption{MNIST}
\label{subfig:full-rec-mnist}
\end{subfigure}
\begin{subfigure}{\linewidth}
\hspace{-0.7cm}
\begin{minipage}[m]{0.85\linewidth}
    {\includegraphics[width=\linewidth]{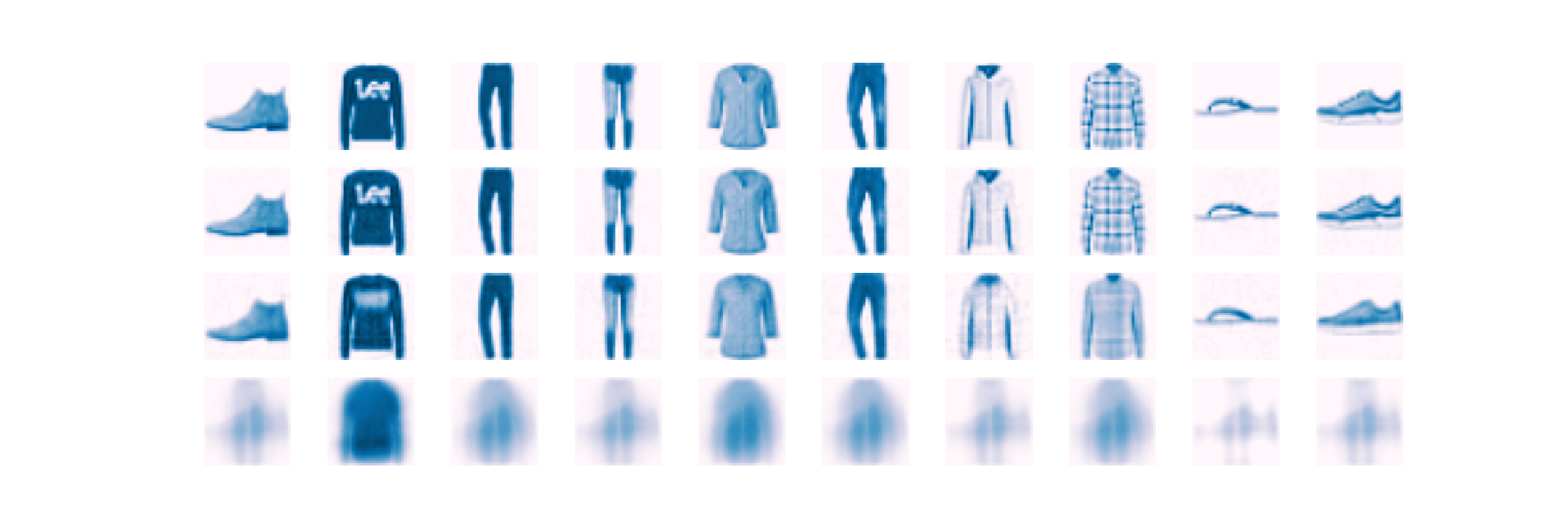}}
\end{minipage}
\hspace{-0.7cm}
\begin{minipage}[m][0.25\linewidth][l]{0.2\linewidth}
    \vfill
    \tiny{original}\vfill
    \tiny{1024\\(SSIM=0.94)}\vfill
    \tiny{64\\(SSIM=0.79)}\vfill
    \tiny{1\\(SSIM=0.22)}\vfill
\end{minipage}
\caption{FashionMNIST}
\label{subfig:full-rec-fmnist}
\end{subfigure}
\caption{Reconstructed images using the classical neural network trained on the entire datasets; first row - several images of the test set; others are the same images reconstructed by a classical neural network of 1024, 64, 1 neurons in the input layer; the size of all images is $(32 \times 32)$}
\label{fig:reconstructed-all}
\end{figure}

For the classical neural network, we also obtained results for the full version of the dataset, because it does not take as long as in the quantum case. In Fig. \ref{fig:reconstructed-all}, images reconstructed by classical neural networks are presented. NNs reconstruct an image from a different number of measurements in the Hadamard basis. Reducing the number of measurements from 1024 to 64 leads to a slight decrease in the detail of the reconstructed image, and reducing to 1 leads to the reconstruction of an image averaged over several different classes.

The single-pixel imaging method also allows to obtain RGB images \cite{RGB-1, RGB-2, RGB-3}. In order to classify such an image without changes in the network architecture, it is enough to convert it to grayscale in the known way 0.2989 * R + 0.5870 * G + 0.1140 * B. To reconstruct a color image in all cases, we need 3 independently trained neural networks to obtain images in the red, green and blue ranges. However, it should be noted that many potential applications of single-pixel visualization do not use multi-channel images and the visible light range.

\section{Quantum computer performance time}\label{secA2}

To estimate the realistic performance time of a quantum computer, with respect to the number of circuits for gradients calculation according to the parameter-shift rule \cite{parameter-shift}, we can use the following expression for a single dataset element:

\begin{equation*}
    t_{\text{el}} = (D_{\text{1q}} \cdot t_{\text{1q}} + D_{\text{2q}} \cdot t_{\text{2q}}) \cdot (2 \cdot n_{\text{params}}^{(q)} + 1) + C,
\end{equation*}
where $t_{1q}$ and $t_{2q}$ are the experimental times of single- and two-qubit gates respectively, $D_{1q(2q)}$ are the circuit depths (maximum number of corresponding gates for one qubit, that cannot be parallelized in the experiment, which is, with respect to the chosen layer circuit structure (Fig. \ref{fig:ansatz}), the sum of the number of layers and the embedding depth), and $C$ is a constant for setup overheads such as register initialization and measurement time and other delays. 

The total time for one epoch is as follows:
\begin{equation*}
    t_{\text{total}} = t_{\text{el}} \cdot N_{\text{shots}} \cdot N_{\text{dataset}}, 
\end{equation*}
where  $N_{\text{shots}}$ is the number of each quantum circuit run to get statistics for the expected value; $N_{\text{dataset}}$ is the number of elements in the dataset.

\end{appendices}

\bibliography{bibliography}

\end{document}